\documentclass[twocolumn,showkeys,showpacs,preprintnumbers,prd,superscriptaddress,nofootinbib]{revtex4-1}
\usepackage{graphicx}
\usepackage{epsf}
\usepackage{bm}
\usepackage{amsmath}
\usepackage{amsfonts}
\usepackage{amssymb}
\usepackage{epstopdf}
\usepackage{natbib}
\usepackage{hyperref}
\usepackage{color}
\usepackage{verbatim}
\usepackage{multirow}
\usepackage{bm}
\usepackage{hyperref}

\begin{document}

\title{Dark sector interaction and the supernova absolute magnitude tension}

\author{Rafael C. Nunes}
\email{rafadcnunes@gmail.com}
\affiliation{Divis\~{a}o de Astrof\'{i}sica, Instituto Nacional de Pesquisas Espaciais, Avenida dos Astronautas 1758, S\~{a}o Jos\'{e} dos Campos, 12227-010, S\~{a}o Paulo, Brazil}

\author{Eleonora Di Valentino}
\email{eleonora.di-valentino@durham.ac.uk}
\affiliation{Institute for Particle Physics Phenomenology, Department of Physics, Durham University, Durham DH1 3LE, UK.}

\begin{abstract}
It has been intensively discussed if modifications in the dynamics of the Universe at late times is able or not to solve the $H_0$ tension. On the other hand, it has also been argued that the $H_0$ tension is actually a tension on the supernova absolute magnitude $M_B$. In this work, we robustly constraint $M_B$ using Pantheon Supernovae Ia (SN) sample, Baryon Acoustic Oscillations (BAO), and Big Bang Nucleosynthesis (BBN) data, and assess the $M_B$ tension by comparing three theoretical models, namely the standard $\Lambda$CDM, the $w$CDM and a non-gravitational interaction (IDE) between dark energy (DE) and dark matter (DM). We find that the IDE model can solve the $M_B$ tension with a coupling different from zero at 95\% CL, confirming the results obtained using a $H_0$ prior.
\end{abstract}

\keywords{}

\pacs{}

\maketitle

\section{Introduction}
\label{sec:introduction}

The standard $\Lambda$-Cold Dark Matter ($\Lambda$CDM) scenario provides a wonderful fit to astronomical observations carried out over the past two decades. On the other hand, recently some tensions and anomalies became statistically significant while analyzing different data sets, placing the $\Lambda$CDM cosmology in a crossroad. The most discussed and statistically significant tension in the literature is in the estimation of the Hubble constant, $H_0$, between the Cosmic Microwave Background (CMB) and the direct local distance ladder measurements. Assuming the $\Lambda$CDM scenario, Planck-CMB data analysis provides $H_0=67.4 \pm 0.5$ km s$^{-1}$Mpc$^{-1}$ \cite{Aghanim:2018eyx}, which is in $4.4\sigma$ tension with the local measurement $H_0 = 74.03 \pm 1.42$ km s$^{-1}$Mpc$^{-1}$~\cite{Riess:2019cxk} by the SH0ES team, or $4.2\sigma$ considering the latest one $H_0 = 73.2 \pm 1.3$ km s$^{-1}$Mpc$^{-1}$~\cite{Riess:2020fzl}.
Additionally, many other late time measurements are in agreement with a higher value for the Hubble constant (see the discussion in~\cite{Verde:2019ivm,Riess:2019qba,DiValentino:2020vnx} and in tension with the Planck-CMB estimate, as for example the Megamaser Cosmology Project~\cite{Pesce:2020xfe} that gives $H_0 = 73.9 \pm 3.0{\rm \,km\,s^{-1}\,Mpc^{-1}}$, or using the Surface Brightness Fluctuations~\cite{Blakeslee:2021rqi} that find $H_0=73.3 \pm 0.7 \pm 2.4 {\rm \,km\,s^{-1}\,Mpc^{-1}}$. The lower value of $H_0$ inferred from the Planck-CMB data is instead in very good agreement with Baryon Acoustic Oscillations (BAO) + Big Bang Nucleosynthesis (BBN) constraints~\cite{Alam_2021}, and other CMB experiments like ACTPol-DR4~\cite{Aiola:2020azj} or SPT-3G~\cite{Dutcher:2021vtw}. Finally, there are measurements in the middle that can not discriminate between the two values, such as the Tip of the Red Giant Branch~\cite{Freedman:2021ahq} or the astrophysical model dependent measurements based on the strong lensing effect~\cite{Birrer:2020tax}. Motivated by these observational discrepancies, unlikely to disappear completely by introducing multiple and unrelated systematic errors, it has been widely discussed in the literature whether new physics beyond the standard cosmological model can solve the $H_0$ tension (see Refs.~\cite{DiValentino:2020zio,DiValentino:2021izs, perivolaropoulos2021challenges} and reference therein for a review).

The SH0ES team measures the absolute peak magnitude, $M_B$, of Type Ia supernovae (SN), assumed to be standard candles, by calibrating the distances of SN host galaxies to local geometric distance anchors via the Cepheid period luminosity relation. The magnitude $M_B$ is then converted into a value of $H_0$ via the magnitude-redshift relation of the Pantheon SN sample~\cite{Scolnic:2017caz}. Therefore, it has been argued that the $H_0$ tension is actually a tension on the supernova absolute magnitude $M_B$~\cite{efstathiou2021h0,Camarena_2021}, because the SH0ES $H_0$ measurement comes directly from $M_B$ estimates. The CMB constraint on the sound horizon to the SN absolute magnitude $M_B$ using the parametric-free inverse distance ladder predicts $M_B=-19.401 \pm 0.027$ mag~\cite{Camarena_2020}, while the the SN measurements from SH0ES corresponds to $M_B= -19.244 \pm 0.037$ mag~\cite{Camarena_2021}. These measurements are at $3.4\sigma$ tension. Thus, as argued in~\cite{efstathiou2021h0,Camarena_2021}, rather than explaining the $H_0$ tension one should instead focus on the supernova absolute magnitude tension, because this is what the Cepheid calibrations are designed to measure. For this reason, it has been shown that modifications of the expansion history at late times can not help with the $H_0$ and/or $M_B$ tension, motivating discussions and investigations in this direction~\cite{marra2021rapid, Alestas_2021_phantom, Alestas_2021_deforming, adil2021late, zhou2021late, Felice_2020, De_Felice_2021_H0}. On the other hand, other possibilities, including early time solutions that introduce new physics prior to recombination, also has been considered as good alternatives to solve the $H_0$ tension (see~\cite{Poulin_2019,M_rtsell_2018,Niedermann_2021,Ye_2020,vagnozzi2021consistency,Chudaykin_2020,D_Amico_2021,Seto:2021xua,Blinov:2020uvz,Hryczuk:2020jhi,DiValentino:2017oaw,Braglia:2020bym,Lin:2020jcb,Smith:2020rxx,Murgia:2020ryi,Keeley:2020rmo,Velten:2021cqj, krishnan2021does,Vagnozzi:2021tjv,delaMacorra:2021hoh} and reference therein), even if they are not completely successful~\cite{Arendse:2019hev,Lin:2021sfs}.

In particular, alternative scenarios involving a non-gravitational interaction between the two main dark species of our Universe, namely, dark matter (DM) and dark energy (DE), have been intensively studied as a possibility to resolve the current cosmological $H_0$ tension (see for example~\cite{Kumar:2016zpg, Kumar:2017dnp, DiValentino:2017iww, Yang:2018ubt, Yang:2018euj, Yang:2019uzo, Kumar_2019, Pan:2019gop, Pan:2019jqh, DiValentino:2019ffd, Di_Valentino_2020_ide03, DiValentino:2020leo, DiValentino:2020kpf, Gomez-Valent:2020mqn, Yang:2019uog, Lucca:2020zjb, Martinelli:2019dau, Yang:2020uga, Yao:2020hkw, Pan:2020bur, DiValentino:2020vnx, Yao:2020pji, Amirhashchi:2020qep, yang20212021h0, Gao:2021xnk, lucca2021dark, kumar2021remedy,Yang:2021oxc,Lucca:2021eqy,halder2021investigating}
for a short list and reference therein).
This class of models has been listed as an example of "not working" solution, i.e. able to solve the $H_0$ tension, but unable to alleviate the $M_B$ tension. Therefore, we aim with this paper to check if modifications to the late time evolution involving a non-gravitational coupling different from zero in the dark sector, can bring the $M_B$ measured from distant supernovae of the Pantheon sample and calibrated by BAO + BBN into agreement with the value obtained calibrating the local SN with the Cepheid measurements. We are interested in the background only analysis, so we don't include CMB data in this paper. But, without loss of generality, using the mentioned data, we will get good accuracy in our constraints. Our main conclusion is that an Interacting DE-DM model (IDE) with a dark coupling different from zero is able to solve the $H_0-M_B$ tension, and it is in agreement with the Pantheon SN sample.

This paper is structured as follows.  
In Section~\ref{sec:theory} we introduce the Interacting DE-DM model considered.
In Section~\ref{sec:data}, we present the data sets and methodology used in this work. In Section~\ref{sec:results}, we discuss the main results of our analysis. In Section~\ref{sec:conclusions}, we outline our final considerations and perspectives.

\section{Theoretical framework}
\label{sec:theory}

In this Section we introduce the Interacting DE-DM model studied here.
In a homogeneous and isotropic Universe, the dark interaction is quantified as

\begin{eqnarray}
\label{DE_DM_1}
\nabla_{\mu}T_{i}^{\mu \nu }=Q_{i}^{\nu}\,, \quad \sum\limits_{\mathrm{i}}{%
Q_{i}^{\mu }}=0~,
\end{eqnarray}
where the index $i$ runs over DM and DE. The four-vector $Q_{i}^{\mu}$ governs the interaction. We can assume that $Q_{i}^{\mu}$ is given by 
\begin{eqnarray}
Q_{i}^{\mu}=(Q_{i}+\delta Q_{i})u^{\mu}+a^{-1}(0,\partial^{\mu}f_{i}), \end{eqnarray}
where $u^{\mu}$ is the velocity four-vector and $Q_i$ is the  background energy transfer. Let us note that from now on we shall use the notation $Q_i \equiv Q$.  The symbol $f_i$ refers to the momentum transfer potential. In the Friedmann-Lema\^{i}tre-Robertson-Walker (FLRW) background, one can write down the conservation equations of the DM and DE densities as 

\begin{eqnarray}
&&\dot{\rho}_c +3 H \rho_c  = Q~,\label{cont1}\\
&&\dot{\rho}_x +3 H (1 + w)\rho_x= - Q~,\label{cont2}
\end{eqnarray}
where $H =\dot{a}/a$ is the expansion rate of the Universe and $w$ is the equation of state (EoS) of DE. 

In the present work, we consider a very well known parametric form of the interaction function $Q$, namely, $Q = \mathcal{H} \xi \rho_x$, where $\xi$ is the coupling parameter between the dark components. From the sign of $\xi$, one can identify the direction of the energy flow between the dark sectors. The condition $\xi < 0$ corresponds to the energy flow from DM to DE, and $\xi > 0$ represents the opposite scenario. The functional form $Q = \mathcal{H} \xi \rho_x$ can avoid the instabilities in the perturbations at early times on the dark sector species, although it will not be necessary to consider the evolution of the perturbations in this work. It should be noted that the inclusion of the global factor $\mathcal{H}$ into the interaction function $Q$ was motivated to quantify a possible global interaction through cosmic history. Although, as already argued by the authors~\cite{V_liviita_2008}, the interaction in the dark sectors should depend on the local quantities too. Nevertheless, as the present interaction function is widely studied in the literature, we aim to revisit this model.

On the other hand, the choice of the interaction function is not unique and it is very difficult to provide a specific functional form since the nature/properties of both dark components is completely unknown at present. From the observational perspective, we do not find any strong signal which could reveal the nature of the dark components, and hence, we believe that the nature of the interaction function will probably remain unknown for the next decade(s). We can only approximate the interaction function $Q$ through theoretical arguments and the consistency with observational data. However, it has been argued by many investigators that the interaction between the dark components may appear from some effective field theory~\cite{Gleyzes_2015}, disformal coupling~\cite{de_Bruck_2015}, axion monodromies~\cite{D_Amico_2016}, varying dark matter mass and fundamental constants~\cite{Marsh_2017} and Horndeski theories~\cite{Kase_2020}. Thus, one can see that some viable formalism can be given for the interacting dark energy theory. Motivated to check the strength of this dark coupling and its observational viability, let us assume in the present work the simple parametric function $Q = \mathcal{H} \xi \rho_x$ to quantify such dark interaction.

\begin{table*}
    \centering
    \caption{Constraints at 95\% CL for the four models under consideration in this work for the Pantheon + BAO + BBN joint analysis. The parameter $H_{\rm 0}$ is measured in units of km s${}^{-1}$ Mpc${}^{-1}$ and $M_B$ in units of mag. The last column represents the degree of tension with $M_B$ from SH0ES.}
    \label{tab:main_results_1}
    \begin{tabular}{ccccccc}
        \hline
        \hline
		Model & $M_B$ & $H_0$ &$\Omega_m$ & $w$ & $\xi$ & Tension \\ 
		\hline
		$\Lambda$CDM & $-19.393^{+0.042}_{-0.040}$ & $68.5^{+1.2}_{-1.1}$ &  $0.302^{+0.025}_{-0.025}$ & -1 & 0 & 3.4$\sigma$\\ 
		$w$CDM & $-19.376^{+0.095}_{-0.11}$ & $69.2^{+3.3}_{-3.7}$ & $0.306^{+0.032}_{-0.034} $ & $-1.02^{+0.11}_{-0.11}$ & 0 & 2.1$\sigma$ \\ 
		IDE & $-19.385^{+0.094}_{-0.089}$ & $69.1^{+3.1}_{-2.8}$ & $0.274^{+0.044}_{-0.050}$ & -0.999 &$> -0.35$  & 2.4$\sigma$ \\
		\hline
    \end{tabular}
\end{table*}

\begin{table*}
    \centering
    \caption{Constraints at 95\% CL for the four models under consideration in this work for the Pantheon + BAO + BBN + $M_B$ joint analysis. The parameter $H_{\rm 0}$ is measured in units of km s${}^{-1}$ Mpc${}^{-1}$ and $M_B$ in units of mag. The last column represents the degree of tension with $M_B$ from SH0ES.}
    \label{tab:main_results+MB}
    \begin{tabular}{ccccccc}
        \hline
        \hline
		Model & $M_B$ & $H_0$ &$\Omega_m$ & $w$ &$\xi$ & Tension \\ 
		\hline
		$\Lambda$CDM & $-19.340^{+0.047}_{-0.046}$ & $70.0^{+1.4}_{-1.4}$ &  $0.314^{+0.024}_{-0.025}$ & -1 & 0 & 2.1$\sigma$\\ 
		$w$CDM & $-19.285^{+0.049}_{-0.060} $ & $72.4^{+1.9}_{-2.2} $ & $0.323^{+0.028}_{-0.025}$ & $-1.10^{+0.09}_{-0.10}$ & 0 & 1.0$\sigma$ \\ 
		IDE & $-19.288^{+0.064}_{-0.063}$ & $72.3^{+2.4}_{-2.3}$ & $0.256^{+0.061}_{-0.065}$ & -0.999& $-0.31^{+0.27}_{-0.28}$  & 0.9$\sigma$ \\
		\hline
    \end{tabular}
\end{table*}

\begin{figure*}
\begin{center}
\includegraphics[width=3.in]{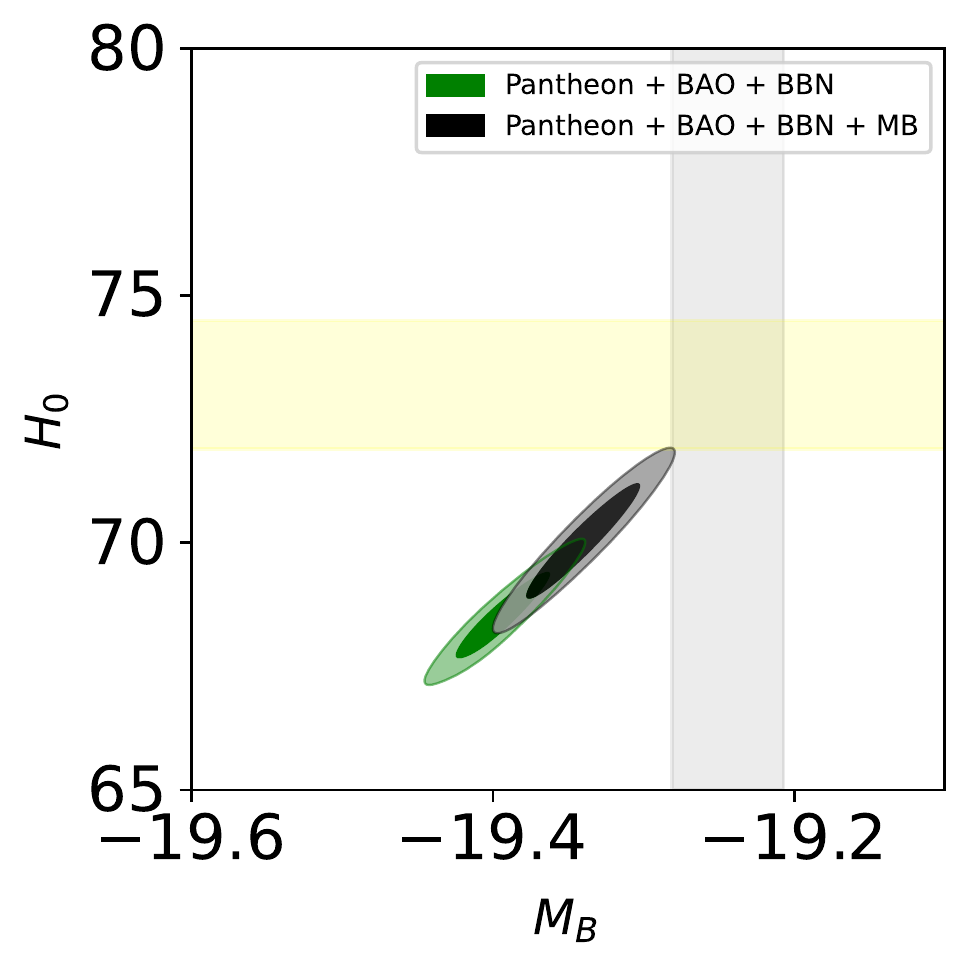} \,\,\,\,\,\,
\includegraphics[width=3.in]{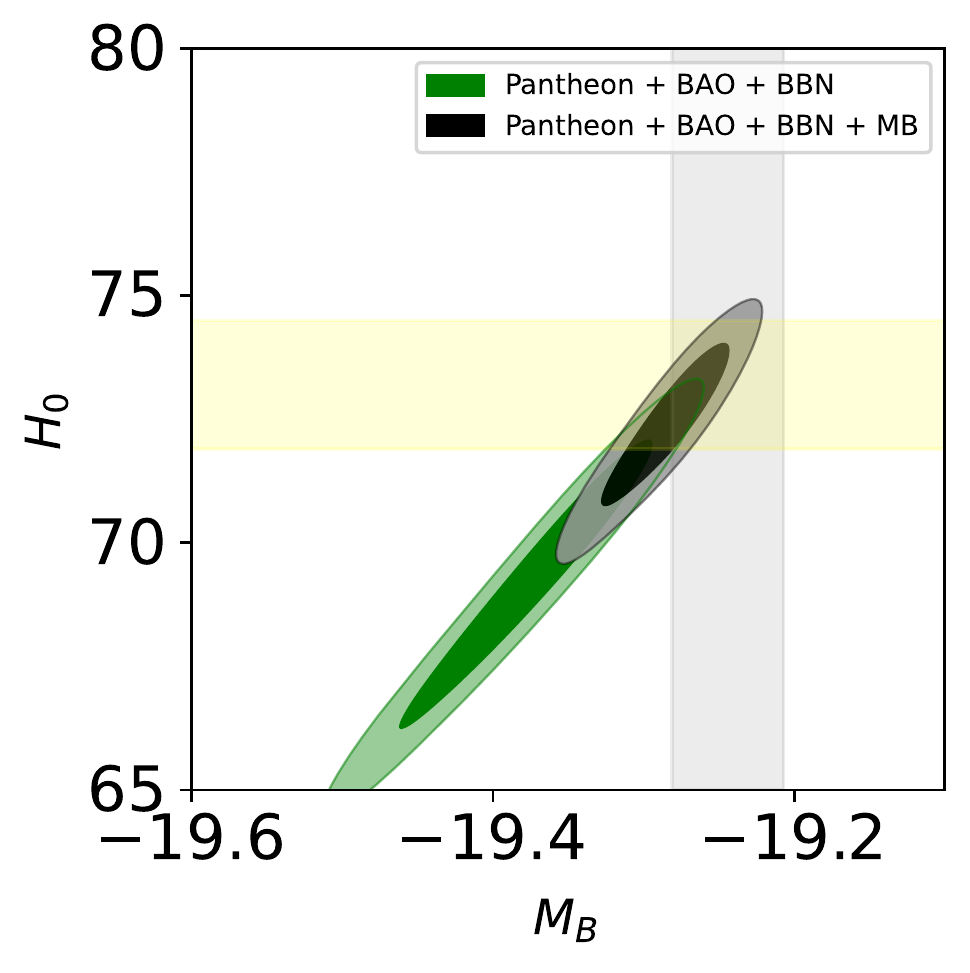}
\includegraphics[width=3.in]{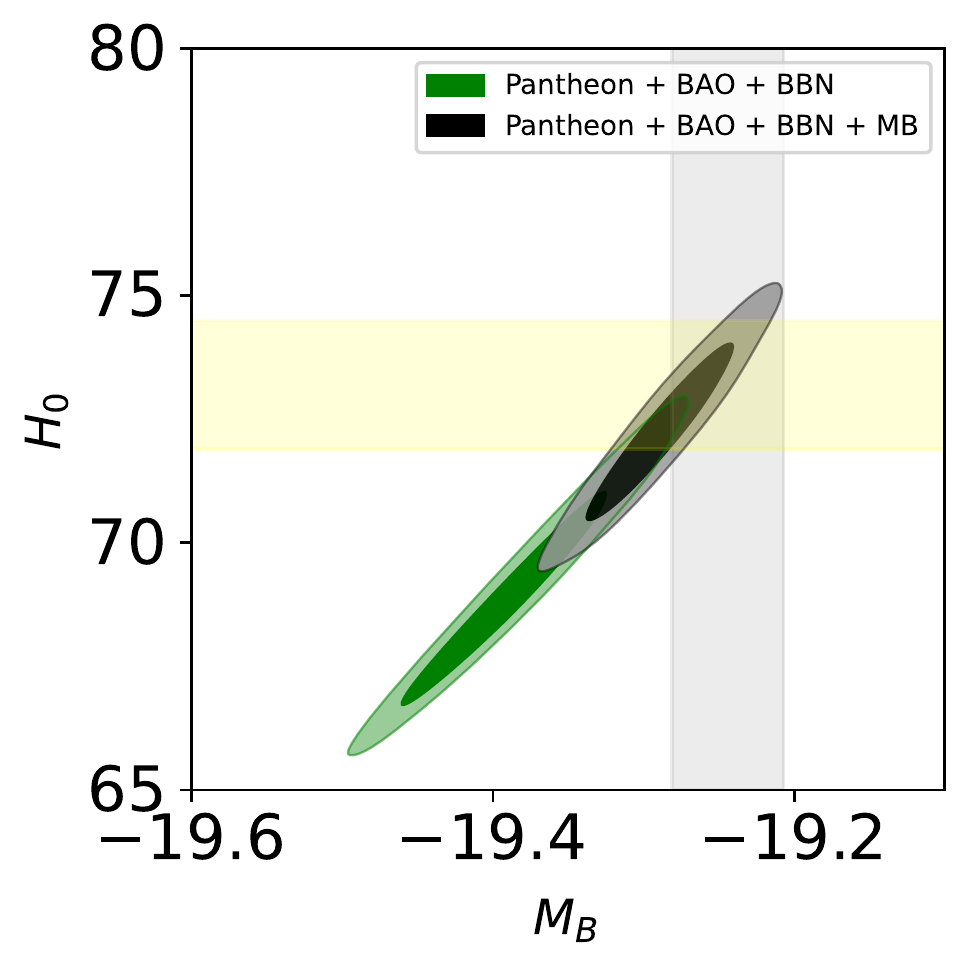}
\caption{Two-dimensional joint posterior distributions in the $M_B$-$H_0$ plane, with the corresponding 68\%~CL and 95\%~CL contours, obtained for Pantheon + BAO + BBN and Pantheon + BAO + BBN + $M_B$ joint analyses within the $\Lambda$CDM (top left panel), $w$CDM (top right panel), and IDE (bottom panel) models. The vertical grey band denotes the $68\%$~CL interval of the SN absolute magnitude $M_B$ based on the SH0ES measurements. The horizontal yellow band denotes the $68\%$~CL interval of the $H_0$ parameter based on the SH0ES measurements.}
\label{M_H0_plot_1}
\end{center}
\end{figure*}

\section{Datasets and Methodology}
\label{sec:data}

In this Section, we present the data sets and methodology used to obtain the observational constraints on the model parameters by performing Bayesian Monte Carlo Markov Chain (MCMC) analysis. In order to constrain the parameters, we use the following data sets:

\begin{itemize}

\item Type Ia Supernovae distance moduli measurements from the \textit{Pantheon} sample~\cite{Scolnic:2017caz}. These measurements constrain the uncalibrated luminosity distance $H_0d_L(z)$, or in other words the slope of the late-time expansion rate (which in turn constrains $\Omega_m$). We refer to this dataset as \textit{Pantheon}. For a SN at redshift $z$, the theoretical apparent magnitude $m_B$ is given by

\begin{eqnarray}
m_B = 5 \log_{10} \Big[ \frac{d_L(z)}{1 Mpc} \Big] + 25 + M_B,
\end{eqnarray}
where $M_B$ is the absolute magnitude. The distance modulus is then written as $\mu(z) = m_B - M_B$.

\item Baryon Acoustic Oscillations (BAO) distance and expansion rate measurements from the 6dFGS~\cite{Beutler:2011hx}, SDSS-DR7 MGS~\cite{Ross:2014qpa}, BOSS DR12~\cite{Alam:2016hwk} galaxy surveys, as well as from eBOSS DR14 Lyman-$\alpha$ (Ly$\alpha$) absorption~\cite{Agathe:2019vsu} and Ly$\alpha$-quasars cross-correlation~\cite{Blomqvist:2019rah}. These consist of isotropic BAO measurements of $D_V(z)/r_d$ (with $D_V(z)$ and $r_d$ the spherically averaged volume distance, and sound horizon at baryon drag respectively) for 6dFGS and MGS, and anisotropic BAO measurements of $D_M(z)/r_d$ and $D_H(z)/r_d$ (with $D_M(z)$ the comoving angular diameter distance and $D_H(z)=c/H(z)$ the Hubble distance)for BOSS DR12, eBOSS DR14 Ly$\alpha$, and eBOSS DR14 Ly$\alpha$-quasars cross-correlation. 

\item The state-of-the-art assumptions on Big
Bang Nucleosynthesis (BBN). The BBN data consist of measurements of the primordial abundances of helium, $Y_P$, from~\cite{Aver_2015}, and the deuterium measurement, $y_{DP} = 10^5 n_D/n_H$, obtained in~\cite{Cooke_2018}. This BBN likelihood is sensitive to the constraints of the physical baryon density $\omega_b \equiv \Omega_bh^2$ and the effective number of neutrino species $N_{\rm eff}$. Let us fix $N_{\rm eff} = 3.046$ in this present work.

\item A gaussian prior on $M_B= -19.244 \pm 0.037$ mag~\cite{Camarena_2021}, corresponding to the SN measurements from SH0ES.

\end{itemize}

We first consider the standard $\Lambda$CDM model, spanned by the following parameters: the Hubble constant $H_0$, the physical baryon density $\omega_b \equiv \Omega_bh^2$, the physical cold dark matter density $\omega_c \equiv \Omega_ch^2$, and the absolute peak magnitude, $M_B$ of SN. The matter density parameter today $\Omega_m$ is given by $\Omega_m = (\omega_b+\omega_c)/h^2$. Then, we consider two 1-parameter extensions of the previous model. Firstly, we introduce a dark energy equation of state free to vary $w$ and we call this model $w$CDM. Secondly, we consider a dark coupling between DM and DE with a single free parameter $\xi$,and we refer to this extended scenario as the IDE model.

It should be noticed here that in order to avoid the gravitational instabilities present when $w =-1$~\cite{Valiviita:2008iv,He:2008si}, we fix the dark energy equation of state to $w = -0.999$, as previously done in~\cite{Salvatelli:2013wra,DiValentino:2019ffd,DiValentino:2019jae}, without affecting the results, as shown by using simulated data in~\cite{DiValentino:2020leo}. Moreover, in order to avoid early-time instabilities~\cite{Valiviita:2008iv,Gavela:2009cy,Clemson:2011an,Gavela:2010tm,He:2008si,Jackson_2009}, we also assume $\xi < 0$.

We use the Metropolis - Hastings mode in \texttt{CLASS}+\texttt{MontePython} code~\cite{Blas:2011rf,Audren:2012wb,Brinckmann:2018cvx} to derive the constraints on cosmological parameters from the data sets described above in this section~\ref{sec:data}, ensuring a Gelman - Rubin convergence criterion of $R - 1 < 10^{-3}$~\cite{Gelman:1992zz}.

\section{Results}
\label{sec:results}

In this work, we consider two different data combinations, namely, Pantheon + BAO + BBN and Pantheon + BAO + BBN +$M_B$ applied on three different cosmological models, namely, $\Lambda$CDM, $w$CDM, and IDE models.

The first joint analysis, shown in Table \ref{tab:main_results_1}, is chosen to break any possible degeneracy in the $H_0-\Omega_m$ plane, because we know that BAO + BBN can constrain these parameters very well. Also, several authors have pointed out the ability of combined BAO and BBN data to probe the background cosmological history, independently of CMB data. As a second step for the analysis, we will add a $M_B$ gaussian prior corresponding to the SN measurements from SH0ES, and we will analyse Pantheon + BAO + BBN + $M_B$, showing the results in Table~\ref{tab:main_results+MB}. 
In this work, we will analyze our baseline regardless of Planck-CMB data and then discuss the potential of the model of solving and/or alleviating the $M_B$ tension. 

Fig~\ref{M_H0_plot_1} show the 2D joint posterior distributions at 68\%~CL and 95\%~CL in the $M_B$-$H_0$ plane, obtained from Pantheon + BAO + BBN and Pantheon + BAO + BBN + $M_B$ for the three scenarios considered in this work. In particular we compare $\Lambda$CDM in the top left panel of Fig~\ref{M_H0_plot_1}, the $w$CDM framework in the top right panel of Fig~\ref{M_H0_plot_1}, and the IDE scenario in the bottom panel of Fig~\ref{M_H0_plot_1}.

\begin{figure*}
\begin{center}
\includegraphics[width=3.in]{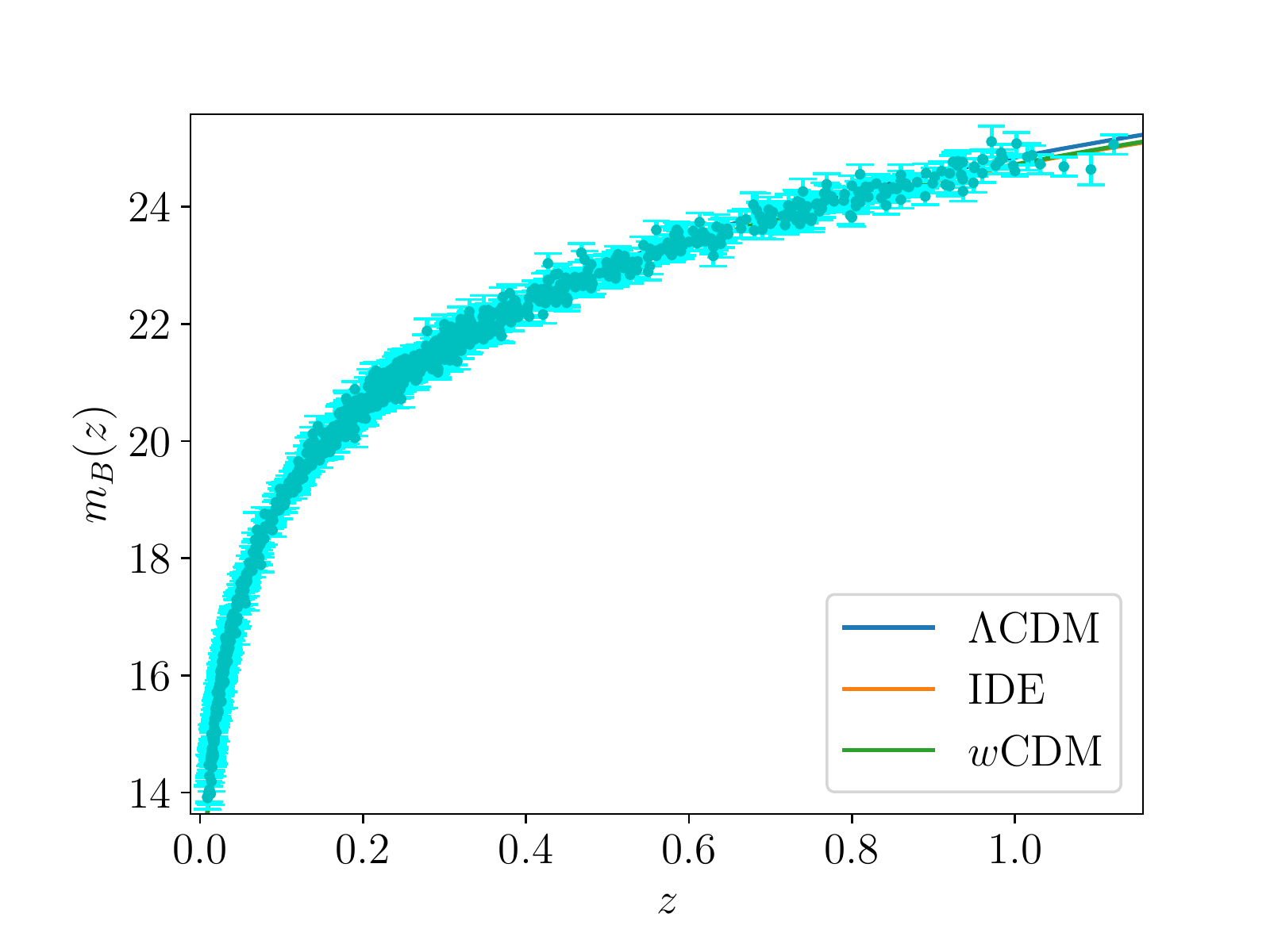} \,\,\,\,\,\,
\includegraphics[width=3.in]{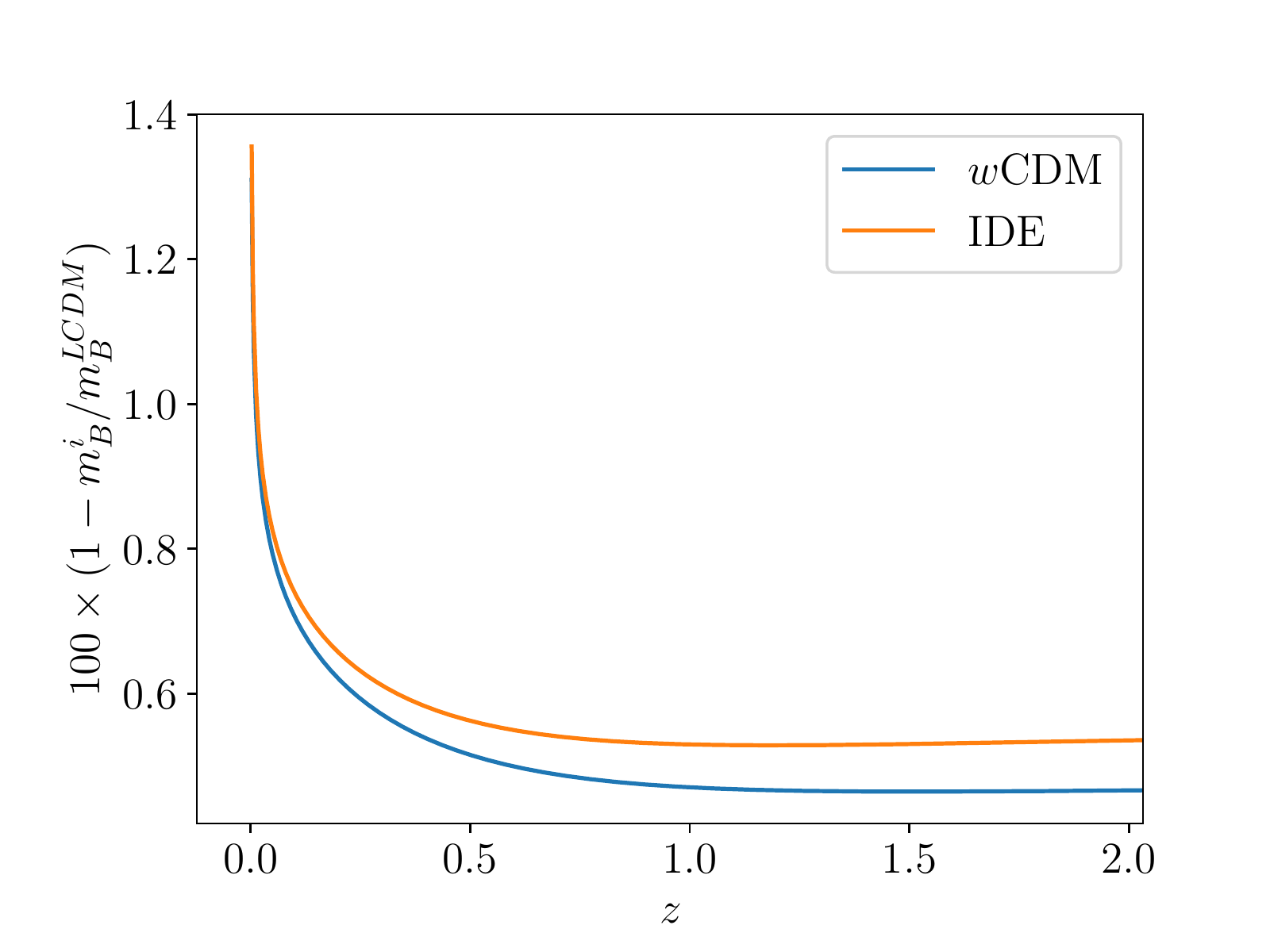} 
\caption{Left panel: Best fit values for the $\Lambda$CDM, $w$CDM and IDE models obtained by Pantheon + BAO + BBN + $M_B$ compared against the magnitude-redshift relation of the Pantheon SN sample in the redshift range $0< z <1$.  Right panel: $100 \times (1-m_B^{i}/m_B^{\Lambda CDM})$, where $i$ run on the $w$CDM and IDE models from the difference of the best fit values between $\Lambda$CDM and $w$CDM or IDE of the left plot}.
\label{SN_plot_2}
\end{center}
\end{figure*}

Once we have obtained constraints on the above cosmological parameters, we can quantify the level of concordance or discordance (if any) on $M_B$ between different data-set $i$ and $j$ using a simple 1D tension metric as follows: 

\begin{equation}
T_{M_B} \equiv \frac{M_{B,i}-M_{B,j}}{\sqrt{\sigma^2_{M_{B,i}}+\sigma^2_{M_{B,j}}}}\,,
\label{eq:1D_estimator}
\end{equation}
where the value of $T_{M_B}$ can directly be interpreted as the level of tension in equivalent Gaussian $\sigma$.

The first thing to notice is that, taking as a reference the SN measurements from SH0ES team obtained by Ref.~\cite{Camarena_2021}, $M_B= -19.244 \pm 0.037$, our constraints on $M_B$ are in tension above $3\sigma$ for the $\Lambda$CDM model.

On the contrary, both the $w$CDM and IDE models can reduce the $M_B$ tension below $3\sigma$. This shows that a late time modification such as a phantom dark energy equation of state can alleviated the $M_B$ tension as suggested when a $H_0$ prior is used.

Regarding the IDE scenario instead, the constraining power of the dataset combinations we are using is not enough to rule out the value of the coupling found from the works in the literature to alleviate the Hubble tension~\cite{Kumar:2016zpg, Kumar:2017dnp, DiValentino:2017iww, Yang:2018ubt, Yang:2018euj, Yang:2019uzo, Kumar_2019, Pan:2019gop, Pan:2019jqh, DiValentino:2019ffd, Di_Valentino_2020_ide03, DiValentino:2020leo, DiValentino:2020kpf, Gomez-Valent:2020mqn, Yang:2019uog, Lucca:2020zjb, Martinelli:2019dau, Yang:2020uga, Yao:2020hkw, Pan:2020bur, DiValentino:2020vnx, Yao:2020pji, Amirhashchi:2020qep, yang20212021h0, Gao:2021xnk, lucca2021dark, kumar2021remedy}. 
Actually, the dark coupling estimate obtained in Table~\ref{tab:main_results_1} has just a lower limit and it is in agreement with both zero and the large coupling needed to solve the $H_0$ tension, within one standard deviation, because of the large error bars. 

Considering the Pantheon + BAO + BBN combination (Table~\ref{tab:main_results_1}), we can see that the introduction of a non null interaction between DM and DE shifts slightly both $H_0$ towards higher values and $M_B$ towards lower values, relaxing the error bars and reducing the tension with SH0ES below $3\sigma$ (see also Fig.~\ref{M_H0_plot_1}, bottom panel).

If we now add a gaussian prior $M_B$ to the previous dataset combination, we obtain the results in Table~\ref{tab:main_results+MB}. Here we can see that, even if in disagreement with the $\Lambda$CDM model at more than $3\sigma$, the addition of the $M_B$ prior can shift its value reducing the tension below $3\sigma$. On the contrary, 
for the $w$CDM model, the combination with the gaussian $M_B$ prior solves the tensions with the SH0ES measurement thanks to the evidence for a phantom dark energy at more than 95\% CL. Moreover,
for the IDE model the $M_B$ prior can be included safely, restoring the complete agreement with the SH0ES data at the price of a negative coupling $\xi=-0.31^{+0.27}_{-0.28}$ at 95\% CL. This result confirms the claim we find in the literature~\cite{Kumar:2016zpg, Kumar:2017dnp, DiValentino:2017iww, Yang:2018ubt, Yang:2018euj, Yang:2019uzo, Kumar_2019, Pan:2019gop, Pan:2019jqh, DiValentino:2019ffd, Di_Valentino_2020_ide03, DiValentino:2020leo, DiValentino:2020kpf, Gomez-Valent:2020mqn, Yang:2019uog, Lucca:2020zjb, Martinelli:2019dau, Yang:2020uga, Yao:2020hkw, Pan:2020bur, DiValentino:2020vnx, Yao:2020pji, Amirhashchi:2020qep, yang20212021h0, Gao:2021xnk, lucca2021dark, kumar2021remedy}, i.e. that a flux of energy from the DM to the DE can solve the Hubble tension, and show that the use of a gaussian prior on $H_0$ or $M_B$ is equivalent for this class of models.

To conclude the analysis, we plot the magnitude-redshift relation for the Pantheon sample in the redshift range $0< z <1$, together with the best fits obtained for Pantheon + BAO + BBN + $M_B$ for a $\Lambda$CDM model (blue line), the IDE model (orange line) and the $w$CDM scenario (green line) in the left panel of Fig~\ref{SN_plot_2}. We show instead the $100 \times (1-m_B^{i}/m_B^{\Lambda CDM})$, where $i$ run on the $w$CDM and IDE models from the difference of the best fit values between $\Lambda$CDM and $w$CDM scenarios (blue line), or $\Lambda$CDM and IDE models (orange line) in the right panel of Fig~\ref{SN_plot_2},  i.e. the percentage difference between the best fits of the models with respect to the $\Lambda$CDM case. We find that in the $z$ range adopted in SH0ES SN data analysis, the difference between the IDE scenario and the $\Lambda$CDM one is $<$ 1.4\%, decreasing up to $< 0.6\%$ at high $z$. For the $w$CDM model, we note practically the same difference at very low $z$ ($<$ 1.35\%), and $< 0.5\%$ at high $z$.

To better quantify the agreement (and/or disagreement) between the models and the Pantheon SN sample, we perform a statistical comparison of the IDE model with the $\Lambda$CDM scenario by using the well-known Akaike information criterion (AIC) \cite{Akaike}.

The AIC is defined through the relation
\begin{equation}
\text{AIC} \equiv -2 \ln  \mathcal{L}_{\rm max} + 2 N,
 \end{equation}
where $ \mathcal{L}_{\rm max}$ is the maximum likelihood function of the model, and $N$ is the total number of free parameters in the model. For the statistical comparison, the AIC difference between the model under study and the reference model is calculated for the joint analysis Pantheon + BAO + BBN + $M_B$. This difference in AIC values can be interpreted as the evidence in favour of the model under study over the reference model. It has been argued in~\cite{Tan_2011} that one model can be preferred with respect to another if the AIC difference between the two models is greater than a threshold value $\Delta_{\rm threshold}$. As a rule of thumb, $\Delta_{\rm threshold} = 5$ can be considered the minimum value to assert a strong support in favour of the model with a smaller AIC value, regardless of the properties of the models under comparison~\cite{Liddle_2007}. 

The result we obtain by computing $AIC^{\Lambda CDM} - AIC^{IDE} = \Delta\text{AIC}=1.54$, which shows that the IDE model can best fit the joint analysis than $\Lambda$CDM model, but the $\Lambda$CDM framework cannot be rule out with high statistical significance. Actually, also a simple $\chi^2$ comparison points in the same direction. We find for the fit of Pantheon + BAO + BBN + $M_B$ a $\chi^2_{min} =1046.74$ for the IDE model, and $\chi^2_{min} = 1050.28$ for the $\Lambda$CDM model, considering the same joint analysis, corresponding to a $\Delta \chi^2_{min} = -3.57$ for one additional degree of freedom. This shows that the IDE model can fit these data better than the $\Lambda$CDM scenario. 

A similar result is obtained for the $w$CDM models, that gives $AIC^{\Lambda CDM} - AIC^{wCDM} = \Delta\text{AIC}=1.52$, improving the fit of the joint analysis with respect to the standard $\Lambda$CDM, even if not in a significant way. Moreover, the $\chi^2_{min} =1046.76$ for the $w$CDM model, with a $\Delta \chi^2_{min} = -3.52$ and only one more degree of freedom.

Therefore, even if the $\Lambda$CDM model fits slightly better the SN data, the best fit for the IDE case, which can solve the Hubble tension with a 95\% CL of evidence for a dark coupling different from zero, cannot be excluded by the Pantheon + BAO + BBN + $M_B$ combination of data.
In other words, for this class of Interacting models, solving the $H_0$ or the $M_B$ tension is equivalent, and using a gaussian prior from SH0ES on the Hubble constant does not bias the results. Actually, the constraints obtained are completely in agreement with the magnitude-redshift relation of the Pantheon SN sample.

\section{Final Remarks}
\label{sec:conclusions}

It has been intensively discussed in~\cite{efstathiou2021h0,Camarena_2021} if modifications in the dynamics of the Universe at late times is able or not to solve the $H_0$ tension once BAO and Pantheon data are taken into account. On the other hand, it has also been argued that the $H_0$ tension is actually a tension on the supernova absolute magnitude $M_B$. 

We investigate this issue analysing three different models, namely the $\Lambda$CDM scenario and two possible late times solutions to the Hubble tension, i.e. a $w$CDM model and Interacting DE-DM model. We analyse these scenarios making use of two possible dataset combinations, that are Pantheon + BAO + BBN and Pantheon + BAO + BBN + $M_B$. We find that the IDE model can alleviate the $M_B$ tension, as well as the $H_0$ disagreement, and a coupling different from zero at more than 95\% CL is actually preferred by the Pantheon + BAO + BBN + $M_B$ dataset combination. The same thing it is true also for a phantom solution with $w<-1$, because it can solve the $M_B$ tension, as does for the $H_0$ tension.

As a final step we investigate if the best fit from an interaction between the DM and the DE for the Pantheon + BAO + BBN + $M_B$ data, than can solve the Hubble tension, is in agreement with the Pantheon data and we find that this model is preferred against the $\Lambda$CDM case.
This result allows us to conclude that applying a SH0ES $H_0$ prior, instead of a $M_B$ prior, does not bias the results for an IDE late time solution. On the contrary, if we repeat the same analysis for a phantom DE equation of state, we do not see an indication for this solution from Pantheon data only, that prefers in any case a cosmological constant.

\begin{acknowledgments}
\noindent  
We thank George Efstathiou, Sunny Vagnozzi, Valerio Marra, Stefano Gariazzo and Olga Mena for their useful comments on the draft.
R.C.N. acknowledges financial support from the Funda\c{c}\~{a}o de Amparo \`{a} Pesquisa do Estado de S\~{a}o Paulo (FAPESP, S\~{a}o Paulo Research Foundation) under the project No. 2018/18036-5.
EDV acknowledges the support of the Addison-Wheeler Fellowship awarded by the Institute of Advanced Study at Durham University.
\end{acknowledgments}

\bibliography{PRD}

\end{document}